\documentclass[aps,pre,preprint,superscriptaddress,amsmath,amssymb]{revtex4-2}

\usepackage{graphicx}
\usepackage{tabularx}
\usepackage{color}
\usepackage{bm}
\usepackage{adjustbox}

\newcolumntype{L}[1]{>{\hsize=#1\hsize\raggedright\arraybackslash}X}%
\newcolumntype{R}[1]{>{\hsize=#1\hsize\raggedleft\arraybackslash}X}%
\newcolumntype{C}[1]{>{\hsize=#1\hsize\centering\arraybackslash}X}%
 
\newcommand{\change}[1]{#1}

\begin{document}


\title{Defect dynamics in growing bacterial colonies}
\author{Rachel Los}
\affiliation{Department of Bionanoscience, Kavli Institute of Nanoscience, Delft University of Technology, Delft, The Netherlands}
\author{Duco \surname{van Holthe tot Echten}}
\affiliation{Department of Bionanoscience, Kavli Institute of Nanoscience, Delft University of Technology, Delft, The Netherlands}
\author{Gerhard Nordemann}
\affiliation{Department of Bionanoscience, Kavli Institute of Nanoscience, Delft University of Technology, Delft, The Netherlands}
\author{Martijn Wehrens}
\affiliation{AMOLF, Amsterdam, The Netherlands}
\author{Sander J. Tans}
\affiliation{Department of Bionanoscience, Kavli Institute of Nanoscience, Delft University of Technology, Delft, The Netherlands}
\affiliation{AMOLF, Amsterdam, The Netherlands}
\author{Timon Idema}
\affiliation{Department of Bionanoscience, Kavli Institute of Nanoscience, Delft University of Technology, Delft, The Netherlands}




\date{\today}

\begin{abstract}
Colonies of rod-shaped bacteria constitute a system of colloidal active matter with nematic properties. As a single initial bacterium multiplies through repeated divisions, the resulting colony quickly loses long-range orientational order, but retains locally ordered domains. At the boundaries of these domains, topological defects emerge, which move around randomly as the colony grows. In both experiments and simulations, we find that these defects are created at a rate that corresponds to the exponential growth of the colony, resulting in a stable defect density. \change{Both this defect density and the colony's correlation length are regulated by the aspect ratio of the rod-shaped particles.}
Moreover, we find that the defect dynamics are well described by a Gamma distribution, which is due to repeated divisions and subsequent re-orientations of the bacteria.
\end{abstract}


\maketitle

\textit{Introduction} \textbf{\textemdash} \change{Active particles continuously generate and dissipate mechanical energy, which they can use to move, grow, or exert forces on each other.} Large collections of active particles can show organized motion on length scales much greater than their individual sizes. These non-equilibrium systems can be found on a large variety of length scales, ranging from microtubule organization, through tissue formation and growing bacterial colonies to schools of fish and flocks of birds~\cite{Toner1998, Sanchez2012, Wensink2012, Mehes2014, VanDrongelen2015, Bechinger2016, Doostmohammadi2016PRL, Ladoux2017, Stenhammar2017, Doostmohammadi2018, Tang2019, Alert2020, Beer2020, Debets2021}. \change{A subclass of these active systems are active nematic liquid crystals, which can be characterized as active elongated particles that are symmetric around their center of mass.} Although originating from different physical principles, these systems show, as in passive liquid crystals, orientational alignment and defect formation~\cite{Doostmohammadi2018}. Defects play a fundamental role in liquid crystals by determining essential system properties, such as their optical and electrical properties. In active materials, these defects are actively created resulting in a completely different organizational structure than found in passive systems~\cite{Doostmohammadi2018, Tang2019}, \change{even though their interactions resemble those of their passive counterparts~\cite{Thijssen2020, Pearce2021}}. \change{Recent studies on topological defects suggest that they govern cell death and extrusions in epithelial tissues~\cite{Saw2017} and control the collective dynamics in neural progenitor cell cultures~\cite{Kawaguchi2017}.}
 
\change{Colonies of rod-shaped bacteria are a prime example of a biological system exhibiting active liquid matter like behavior. While biological and chemical processes dominate colony development in low-density systems~\cite{Waters2005}, colony development at high densities is governed by mechanical interactions~\cite{Cho2007,Boyer2011,Ladoux2017}.} Despite being hardly explored for years, the interest in the influence of mechanical interactions on colony structure has strongly increased~\cite{Doostmohammadi2016PRL, Bonachela2011, Rudge2012, Farrell2013, Smith2017, Beroz2018, DellArciprete2018, Jeckel2019, Warren2019, Copenhagen2021}. By using the framework of liquid crystals, the underlying physical mechanisms of bacterial colony growth, and how characteristics of an individual bacterium affect the colony as a whole, can be studied. These have become essential questions in a variety of fields, including theoretical biology, non-equilibrium physics, engineering of micro-devices and the prevention of bacterial colony formation in the first place. 

\change{When a motile bacterium} attaches to a surface, it can start growing and dividing, and by doing so quickly develops into a large two-dimensional surface-attached colony. Most bacteria live in such surface-attached colonies and derived biofilms~\cite{Garrett2008, Duvernoy2018}, in places ranging from teeth and catheters to pipelines and stones in riverbeds. 
Surface-attached two-dimensional growing bacterial colonies show orientational order over a finite length. As the colony grows, the orientational order within the system changes. Starting from a highly aligned structure, domains of local order develop. The emergence of these domains, just as in passive liquid crystals, indicates the existence of topological defects in the system. As opposed to their passive counterparts however, in active systems these defects are actively created by local energy input in the bulk of the system, \change{which leads to interesting dynamical behavior}~\cite{Doostmohammadi2018,Tang2019}.

\change{As bacterial colonies typically contain hundreds to a few thousands of individuals, the properties of the individual bacteria can affect those of the colony. We therefore use agent-based modeling} to capture the local mechanical interactions between active agents which result in complex behavior on a much larger scale. Moreover, it helps us to easily track defects and their dynamics, since they naturally emerge from local interactions. We find, in experiments and simulations, that these defects are created at a rate that corresponds to the exponential growth of the colony, resulting in a stable defect density, \change{as also found in a continuous model~\cite{DellArciprete2018}}. In addition to advective motion due to colony growth, the defects moreover exhibit super-diffusive behavior with Gamma-distributed step sizes. Both of these observations directly result from the discrete nature of the particles in the colony. While continuum models can qualitatively predict some properties of these colonies like their pressure profile, we show that quantitative agreement requires corrections depending on the average aspect ratio of the particles. Furthermore, this aspect ratio is also the main regulator of various other geometric and topological properties of the system, including the local order and correlation length.

\textit{Model} \textbf{\textemdash} We model our rod-shaped bacteria as spherocylinders: cylinders with spherical caps, where the caps are connected by a spring with finite rest length. We let the bacteria grow over time by extending the rest length of their central spring \change{while keeping the width of the spherocylinders, $D$, constant. This diameter provides us with a natural length scale for our system}. Once the \change{internal spring's rest} length has reached a maximum, the bacterium divides into two identical daughter particles. \change{This is where we introduce a bit of noise to ensure that the daughter particles will not perfectly align and synchronize. } First, to prevent the daughters from perfectly aligning, we give their orientation a small deviation. Second, to prevent all bacteria from dividing at the same time, we assign each daughter \change{a maximum length drawn from a normal distribution around the mean value, and we assign them} a growth rate drawn from a narrow normal distribution around the value of their parent.
\change{We also make our bacteria slightly soft, with overlaps resulting in a repulsive force, which we implement in the same way as done by Storck et al.~\cite{Storck2014} (details given in section I in the Supplemental Material).} The dynamics of the bacteria are overdamped, with all the forces originating from the overlaps and the growth.

As a measure of orientational order, we use the scalar order parameter~$S$ from liquid crystal theory. For a nematic liquid crystal with director $\mathbf{n}$, the scalar order parameter is given by 
$S = \langle 2 (\mathbf{w}_i \cdot \mathbf{n}) - 1 \rangle = \langle \cos (2\theta_i) \rangle$,
where $\mathbf{w}_i$ is the orientation of particle $i$, $\theta_i$ the angle between $\mathbf{w}_i$ and $\mathbf{n}$, and the brackets indicate an ensemble average.

\change{\textit{Experimental system} \textbf{\textemdash} For the experiments, 
\textit{E. coli} microcolonies were grown and imaged under a microscope as previously described~\cite{Wehrens2018}. An extensive description can also be found in the supplemental materials and methods, but in brief, the bacteria were introduced either in a microfluidic device or on a gel pad, and held in place by means of a glass cover slip. The device or pad was then placed under a microscope, and different strains of \textit{E. coli} were grown this way under various conditions (see supplementary methods and Table~S2), whilst images were taken at $1.5 - 2$ min intervals (see Fig.~S2 for example snapshots). The movies were segmented and analyzed using custom MATLAB (MathWorks) scripts from the Tans lab~\cite{Kiviet2014, Wehrens2018} that were originally derived from the \textit{Schnitzcells} package~\cite{Young2012}. From the segmented data, areas, lengths and widths were extracted for all cells to determine growth rates and aspect ratios. Additionally, by fitting ellipsoids to the segmented areas, positions and orientations of all the particles were determined. These were used to calculate the orientational order parameters (average over last fifty frames) and defect dynamics (whole movies) using the same methods that were used for the simulation data.}

\begin{figure}[htp]
	\centering
	\includegraphics[scale=1]{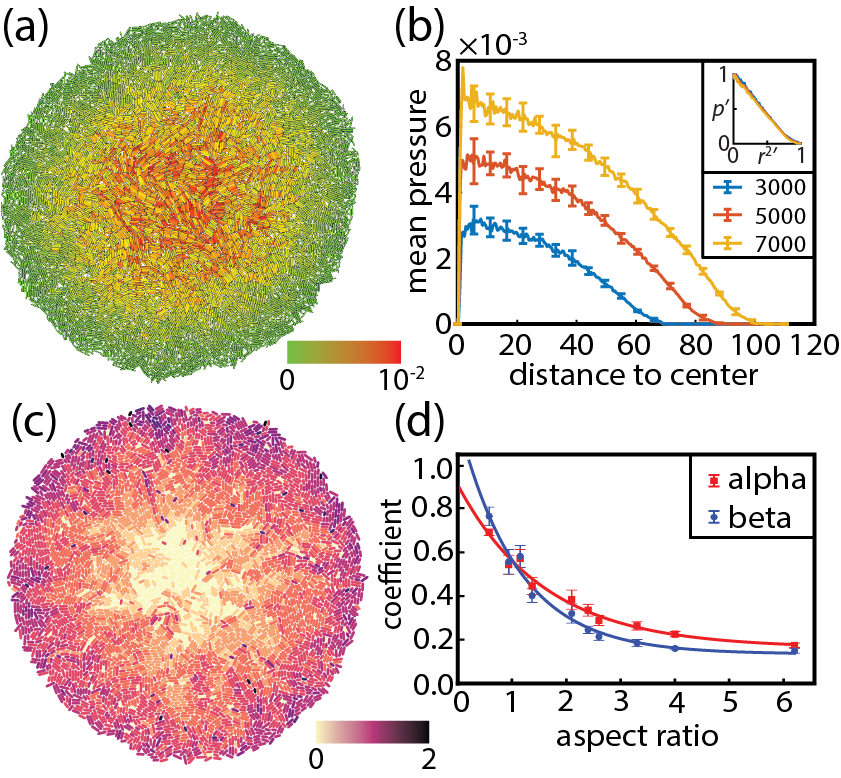}
	\caption{Pressure and velocity profile of simulated colonies. (a) Visualization of the pressure distribution in the colony, ranging from high pressure (red) in the center to zero pressure (green) at the edge. \change{Scale bar ranging from $0$ to $10^{-2}$ in dimensionless units.} (b) Snapshots of the \change{average pressure profile as a function of distance to center measured in units of particle diameter~($D$)} at different  \change{colony sizes}, showing a quadratic decay. Inset: the profiles collapse on a single line for different times as function of distance to center squared \change{measured in units of diameter}. (c) Visualization of the velocity profile in the colony, ranging from low (yellow) in the center to high (purple) at the edge. \change{Scale bar ranging from $0$ to $2$ particle diameters per frame.} (d) Correction factors $\alpha$ (velocity) and $\beta$ (radius) that indicate deviation from naive hydrodynamic model. Continuous lines are fitted functional forms $e^{(\phi-\phi_0)/b}+c$, with $\phi$ the aspect ratio. \change{Error bars in (b) and (d) represent standard deviations over ten runs}.
	}
	\label{fig:stokes}
\end{figure}
\begin{figure}[htp]
\includegraphics[scale=1]{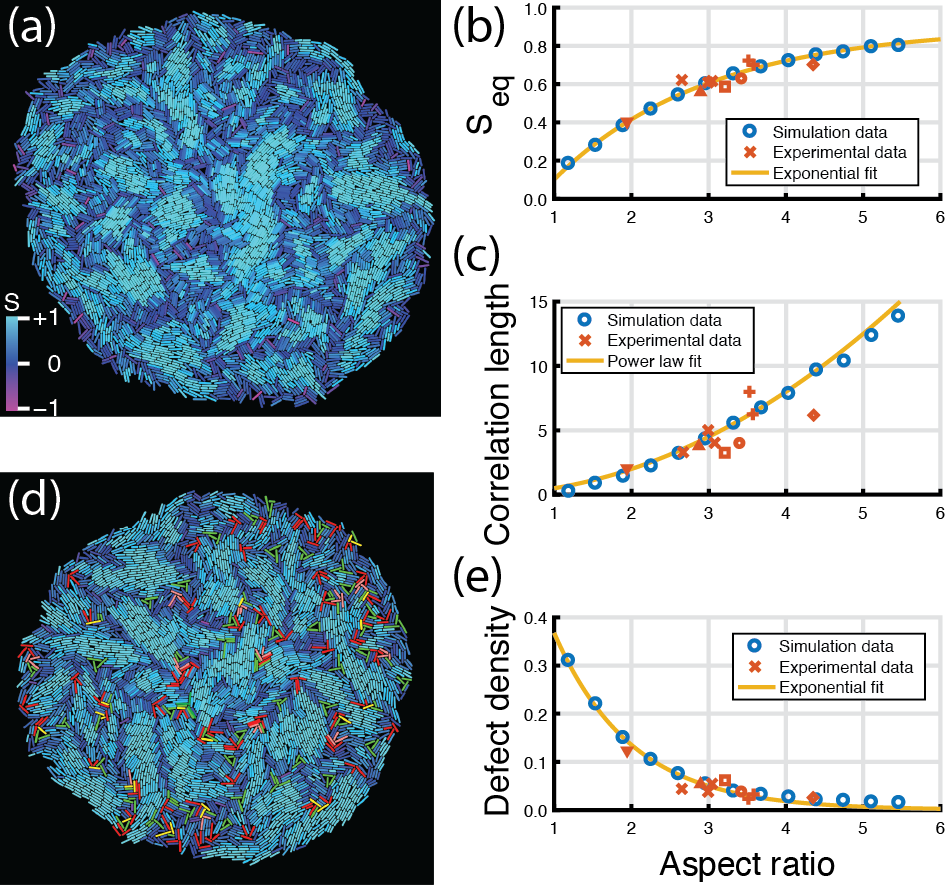}
\caption{Order and defects in bacterial colonies. (a) Visualization of the scalar order parameter $S_\mathrm{p}$ of individual particles for a simulated colony. \change{Diameter of all the particles is equal to 1. }Light blue equals a high value, \change{dark blue zero}, and \change{purple a high negative} value of $S_\mathrm{p}$. (b) Equilibrium value~$S_\mathrm{eq}$ of the mean of $S_\mathrm{p}$ versus the aspect ratio, for simulation (blue circles) and experimental \change{(red symbols)} data. The continuous yellow line is an exponential fit to the simulation data, with fit parameters $a = 0.90$, $b=2.0$, and $\phi_0 = 0.74$. (c) Correlation length of $S_\mathrm{p}$ versus aspect ratio, for simulation \change{(blue circles)} and experimental \change{(red symbols)} data. The continuous yellow line is a power law fit to the simulation data, with fit parameters $a = 0.50$ and $b = 2.0$. \change{A log-log version of this plot can be found in the supplements (Fig.~S9).} (d) Visualization of $+\frac12$ (\change{light/dark} red) and $-\frac12$ (\change{light/dark} green) defects in a snapshot of a simulated colony. \change{Yellow-colored particles are part of both a $+\frac12$ and a $-\frac12$ defect. The diameter of all the particles is equal to $1$. Blue colors as in (a).} (e) Defect density versus aspect ratio for simulation (blue circles) and experimental \change{(red symbols)} data. The continuous yellow line is the exponential relationship $\rho = \exp(-\phi)$.}
\label{fig:colonycharacteristics}
\end{figure}

\textit{Results} \textbf{\textemdash} In both experiments and simulations, we find that over time, a colony that is initiated from a single bacterium becomes round, with distinct regions of aligned particles, as shown in Figs.~\ref{fig:stokes}a, \ref{fig:colonycharacteristics}a, \ref{fig:colonycharacteristics}d, and S2. In our simulations, we find that as the colony grows, the pressure inside builds up (Figs.~\ref{fig:stokes}a and b), even as the particles exhibit a net outward flow (Fig.~\ref{fig:stokes}c). This effect is especially pronounced at higher aspect ratios, because the outward flow is too slow (due to larger drag forces on the longer bacteria) to compensate for the growth. We find a quadratically decaying pressure profile when moving from the center to the edge of the colony. The normalized pressure versus normalized distance to colony center squared collapses to a single line for different times (Fig.~\ref{fig:stokes}b inset). Qualitatively, this quadratically decaying profile can be understood by looking at a simple hydrodynamic model (detailed in section IV in the Supplemental Material). We also find a radially symmetric linear velocity profile, $v= \alpha \frac12 \gamma r$, which only depends on the growth rate, $\gamma$, of individual bacteria, and an aspect ratio correction factor $\alpha$  (Fig.~S11a). Once the colony has become circular, its radius grows exponentially in time as $R=R_0 e^{\beta \frac12 \gamma t}$, in which $\beta$ is also a correction factor (Fig.~S11b). The hydrodynamic model can predict the linear dependence of the velocity on the position, and the exponential dependence of the radius on time, but not the correction factors $\alpha$ and $\beta$. We plot the values of these correction factors as a function of aspect ratio~$\phi$ in Fig.~\ref{fig:stokes}d, together with a fit to an exponential functional form $e^{(\phi-\phi_0)/b}+c$.

We can easily calculate the scalar order parameter~$S$ for an entire colony ($S_\mathrm{col}$) in both the experimental and simulated datasets. Unsurprisingly, we find that it decreases to zero as the number of particles increases (Fig.~S5a). The rate at which~$S$ decreases depends on the average aspect ratio of the particles. More interestingly, we can calculate the value of the scalar order parameter for each individual particle in the colony ($S_\mathrm{p}$),\change{ by} taking the director~$\mathbf{n}$ to lie along the particle, and \change{comparing the average orientation of the} particle's neighbors \change{to it (Fig. ~S3a)}. We then average the value of $S_\mathrm{p}$ over all particles in the colony. This order parameter quickly reaches an equilibrium value as the colony grows (Fig.~S\change{5b}). In Fig.~\ref{fig:colonycharacteristics}a, we color the bacteria by their value of~$S_\mathrm{p}$, and in Fig.~\ref{fig:colonycharacteristics}b we plot the equilibrium value of~$S_\mathrm{eq}$ as a function of the aspect ratio for both our simulated and experimental colonies. The functional form of the relation between the equilibrium order parameter and the aspect ratio follows an exponential:
$S_\mathrm{eq} = a \left[1 - \exp\left(-\frac{\phi-\phi_0}{b}\right) \right]$,
for which we obtain~$a = 0.90$, $b=2.0$, and $\phi_0 = 0.74$ from a fit to the simulation data.

As can be seen in Fig.~\ref{fig:colonycharacteristics}a, the colony contains domains (light blue) of bacteria that are aligned, separated by disordered (dark blue) regions. To characterize the size of the domains, we measure the correlation between orientations as a function of distance. This correlation drops off exponentially (Fig.~S7), which allows us to extract a correlation length, Fig.~\ref{fig:colonycharacteristics}c, both for our simulated and experimental data. The correlation length again depends on the aspect ratio~$\phi$. Their relation follows a simple power law:
$\xi_{\mathrm{S}} = a \phi^b$,
for which we obtain~$a = 0.50$ and $b = 2.0$ from a fit to the simulation data.

In addition to a loss of long-range alignment, colony growth also results in the emergence of orientational defects, places where the director field is not defined. To detect the defects, we use the algorithm of Zapotocky et al.~\cite{Zapotocky1995} \change{(Fig.~S4b)}. We can characterize these defects by counting how many rotations the particles surrounding them make if we go around a defect counter-clockwise. Again in analogy with liquid crystals, we find both $+1/2$ and $-1/2$ defects (meaning that the particles make half a turn when going around the defect), and no $\pm 1$ defects (Fig.~\ref{fig:colonycharacteristics}d). Unsurprisingly, as the colony grows, so does the number of defects. Similar to the orientational order parameter, the defect density~$\rho$ reaches a steady state over time, which also depends strongly on the aspect ratio, following a very simple exponential relationship (Fig.~\ref{fig:colonycharacteristics}e):
$\rho = e^{-\phi}$.

\begin{figure}[htp]
\includegraphics[scale=1]{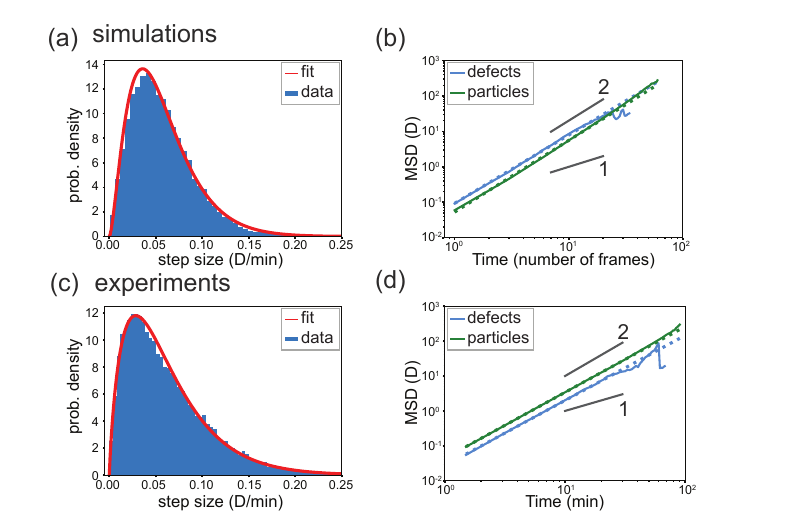}
\caption{Defect dynamics in (a{\&}b) simulations and (c{\&}d) experiments. (a{\&}c) Distribution of defect step sizes, which has a heavy tail and is well described by a Gamma distribution (red lines). Step sizes are measured in units of \change{particle diameter~($D$)}. (b{\&}d) Mean-squared displacements (MSDs) of the defects and particles over time.}
\label{fig:defectdynamics}
\end{figure}

Not only are new defects created (and occasionally destroyed) over time, the position of the defects also changes, due to the growing of the bacteria. Fig.~\ref{fig:defectdynamics} shows \change{the distribution of the step sizes, where step size is defined as the displacement per minute or per frame, and the }mean-squared displacements of the defects in our simulated and experimental systems.
In both systems, we find that the step sizes follow a Gamma distribution (Figs.~\ref{fig:defectdynamics}a and c). \change{For the simulations, we find a peak that is more to the right, which is reflected in a bigger shape parameter of $2.66$, as opposed to $1.86$ for the experiments. Moreover, we see that the simulation data has a slightly narrower distribution with a higher peak, which can be seen in the higher rate parameter of $46.26$ as opposed to $30.15$ for the experiments. We find that the mean-squared displacement of the two types of defects is identical, and that they exhibit the same super-diffusive behavior as the bacteria themselves (Figs.~\ref{fig:defectdynamics}b and d). The slope of the MSD fits for the defects and particles in the simulations are $1.93$ and $2.02$, whereas for the experiments they are $1.87$ and $1.89$, respectively}

\textit{Discussion} \textbf{\textemdash} Bacterial colonies are an example of an active liquid-crystal-like colloidal system. Any system with rod-shaped particles has both positional and orientational degrees of freedom, and can therefore exhibit nematic-like phases. For a passive system, orientational domains may emerge when transitioning from an isotropic into a nematic phase, but as the system equilibrates, these domains will coarsen and eventually cover the whole system. In active liquid crystals however, defect pairs may spontaneously be created~\cite{Sanchez2012,Giomi2013,Keber2014}.

In our case, the system is active only because the bacteria grow and divide. Moreover, instead of transitioning from an isotropic state, the colonies are built due to the repeated division of cells. The orientational order is initially very high, as daughters start their life aligned with each other along the direction of their mother. However, like in the active liquid crystal case, defects and domain boundaries soon emerge as the (grand)daughter cells grow and push each other aside. The overall orientational order therefore quickly drops to zero, but local order remains. Although this latter observation has also been made by others recently~\cite{You2018}, our analysis in terms of the scalar order parameter, correlation length, and defect density allows us to quantify these concepts for the first time. The trends we observe for these quantities as a function of the aspect ratio are described by simple exponential and power law relations. We find that these relations are good fits for both the experimental and simulation data. Our results therefore give us new measures for characterizing the properties of a bacterial colony, and at the same time validate the simulation model used. Moreover, they allow us to compare our results to those of other systems with active particles. In particular, self-propelling particles are known to be able to move at densities well above jamming if they combine their self-propulsion with orientational noise~\cite{Henkes2011,McCusker2019}. In their simulations of large systems of these particles, McCusker et al. found that the orientation of the particles globally orders below a critical noise, and exhibits local order above that noise, allowing for local rearrangements~\cite{McCusker2019}. In our system, the effective orientational noise is high due to the frequent divisions and subsequent rearrangement of daughter cells, but decreases with increasing aspect ratio, leading to a power-law scaling in the correlation length, similar to the dependence of the correlation length on the orientational noise in the self-propelling particle system.

From both our experimental and simulated systems, we find that, on top of the convective motion with the colony, defects exhibit super-diffusive dynamics with Gamma-distributed step sizes. This distribution commonly emerges from diffusive-like behavior with uniform directional changes at exponentially distributed times. These particular random flights are called Pearson-Gamma random walks \cite{LeCaer2011} and are often used to model biological systems, such as planar random flights of micro-organisms  \cite{LeCaer2011,Conolly1987,Stadje1987,Orsingher2007,Codling2008,Kolesnik2008,Beghin2010}. Here we show that they can also emerge from an exponential growth and division process, in combination with mechanical interactions between the daughter cells. The growth of the bacteria leads to the uniform velocity field shown in Fig.~\ref{fig:stokes}c, which is already subtracted from the defect dynamics. Division events on the other hand frequently lead to changes in orientation of the daughter cells, thus making them a likely candidate for causing changes in the direction of a nearby defect. As the distribution of the intervals between these division events is exponential, the observed Gamma distribution of the step sizes of the defects then emerges naturally.

Finally, we note that the aspect ratio in living bacteria can both be controlled by genetic engineering, or change naturally over time by evolution. Hence, we could use our results to design colonies with specific ordering properties, and to predict how effects at the colony scale could influence the evolutionary development of individuals.

\textit{Conclusion} \textbf{\textemdash} We find that, both in experiment and simulations, the average aspect ratio within a freely growing two-dimensional colony governs several essential system properties. First, unsurprisingly, we find that larger aspect ratios push the system away from a continuum description which tacitly assumes isotropic particles. The functional form of the pressure and velocity profiles remains preserved however, with the only change a correction to the growth factor. Second, the equilibrium order parameter increases exponentially with aspect ratio. Third, we find a simple quadratic relation between correlation length and aspect ratio. An increasing equilibrium order parameter and correlation length for larger aspect ratios would imply a decrease in topological defects. Indeed, the \change{equilibrated} defect density decreases exponentially with increasing aspect ratio. Interestingly, we find that these defects show super-diffusive behavior with Gamma-distributed step sizes, which we can understand based on the division dynamics of the bacteria.

\clearpage
\appendix
\setcounter{figure}{0}
\makeatletter
\renewcommand{\thefigure}{S\@arabic\c@figure}
\renewcommand{\thetable}{S\@arabic\c@table}
\makeatother
\section{Model details}
\subsection{Individual-based modeling of sphero-cylindrical particles}
Several agent-based models to simulate the formation of microbial aggregates already exist \cite{Volfson2008,Farrell2013,Balagam2015,Rudge2012,Kreft2001,Ghosh2015}. Our model is based on the particle-spring model developed by Storck \textit{et al.} \cite{Storck2014}, since it is able to control the geometrical properties of the individual cells. Bacteria are modeled by soft, sphero-cylindrical particles that mimic bacilli.

\subsubsection{Single particles}
\begin{figure}[ht]
\begin{center}
\includegraphics[scale=1]{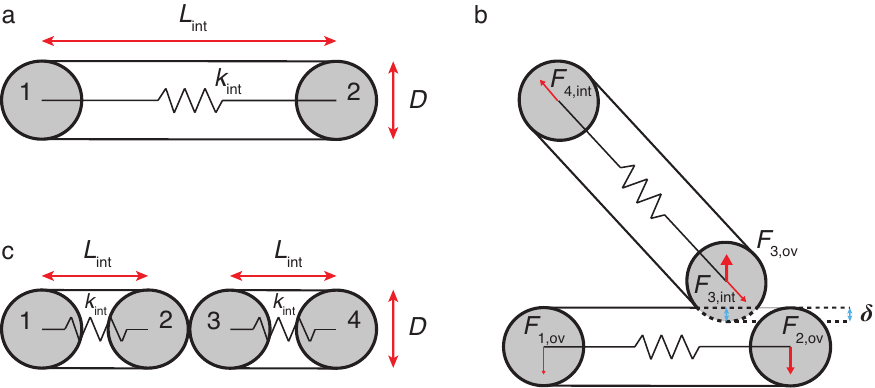}
\end{center}
\caption{Particle model. (a) One particle consists of two disks connected by a spring. (b) When two particles overlap they feel a repulsive force also modeled as a spring. Forces on individual disks are indicated by red force vectors. The width of the force vectors corresponds to the magnitude of the force. (c) When a particle reaches its division length, $L_{\text{int}}^{\text{max}}$, it divides into two daughter particles, each with a length of $L_d = (L_{\text{int}}^{\text{max}} - D)/2$.}
\label{fig:model}
\end{figure}

Following the method of Storck \textit{et al.} \cite{Storck2014}, we define our basic element as a disk $i$ with diameter $D$ and we consider the movement of each disk individually. By connecting two disks with an internal spring that has a length $L_{\text{int}}$ and a spring constant $k_{\text{int}}$ we construct the desired sphero-cylindrical shape of our particles (Fig.~\ref{fig:model}a). To simulate growth, the rest length of the spring $L_{\text{int}}^{\text{rest}}$ is increased every growth step with a constant amount, which is determined by the growth rate $\mu_{\text{int}}$. The disks $i$ and $j$ that make up the particle are then driven apart due to the force applied by the spring, following Hooke's law:
\begin{equation} \label{eq:Fi}
    \bm{F}_{i,\text{int}}\ =\ -k_{\text{int}}\ \frac{\bm{L}_{i,\text{int}}}{L_{i,\text{int}}}\ (L_{i,\text{int}}^{\text{rest}}\ -\ L_{i,\text{int}}),
\end{equation}
where $\bm{L}_{i,\text{int}}$ is the vector from disk $i$ to disk $j$ and is given by
\begin{equation}
    \bm{L}_{i,\text{int}}\ =\ -\bm{L}_{j,\text{int}}\ =\ \bm{r}_j\ -\ \bm{r}_i,
\end{equation}
where $\bm{r}_j$ and $\bm{r}_i$ are the positions of the centers of the disks.
\subsubsection{Collision between particles}
\label{collision}
When two particles collide/overlap they need to push each other away. This collision response is also modeled by a spring with spring constant $k_{\text{ov}}$ and is only active when particles overlap, thus when the distance between particles is smaller than their diameter $D$. A particle needs to interact with other particles as a sphero-cylinder and not as two individual disks. Hence, if two particles overlap, we use a function to compute the shortest distance from line-segment (first particle) to line-segment (second particle) and it returns the size, position and direction of the overlap vector $\bm{d}$ (\cite[Chapter 5]{Ericson2004}). The resulting overlap force is distributed over the two disks $i$ and $j$ of each particle, according to the ratio $\psi_{\text{ov}}$ given by:
\begin{equation}
    \psi_{i,\text{ov}}\ =\ 1-\psi_{j,\text{ov}}\ =\ \frac{|\bm{r}_i - \bm{d}|}{L_{\text{int}}}.
\end{equation}
Summing over the overlapping particles $N_{\text{ov}}$ gives the total overlap force for disk $i$:
\begin{equation}
\label{eq:Fo}
    \bm{F}_{i,\text{ov}}\ =\ \sum_{p \in N_{\text{ov}}} -k_{\text{ov}}\ \frac{\bm{d}_p}{|\bm{d}_p|}(D\ -\ |\bm{d}_p|)\ \psi_{ip,\text{ov}}\ =\ \sum_{p \in N_{\text{ov}}} -k_{\text{ov}}\ \bm{\delta}_{p}\ \psi_{ip,\text{ov}} \qquad \bm{d} < D,\quad \bm{\delta} > 0.
\end{equation}
An example of two overlapping particles and the corresponding forces is illustrated in Fig.~\ref{fig:model}b.

For the scope of this project no further interactions were considered. We tried to keep the parameter space as small as possible and we assumed that the repulsive interactions would dominate the effect on structural order. However, for future studies the model can be extended with other interactions, e.g. adhesion between cells or cell-substrate interactions, which can be done by introducing other springs (as described by Storck \textit{et al.} \cite{Storck2014}).

\subsubsection{Division of particles}
During the development of a biofilm, bacteria are constantly growing and dividing. When a cell reaches its maximal length, it splits into two daughter cells. In the model, the division step is simulated by substituting a particle with two daughter particles, as illustrated in Fig.~\ref{fig:model}c. The maximal length of a particle is set by the maximum length of its internal spring $L_{\text{int}}^{\text{max}}$. In order to fill the same space after the division, the length of the daughter particles $L_d$ is equal to $L_d = (L_{\text{int}}^{\text{max}} - D)/2$, where $L_{\text{int}}^{\text{max}}$ is the maximum length of the mother particle. The daughter particles retain a similar growth rate and orientation as the mother particle (see section \ref{noise}). After division, the internal springs of the daughter cells are at rest.

\subsubsection{Noise}
\label{noise}
To prevent the system from growing in one line, we introduce orientational noise. After division, the orientation of both daughter cells is changed with a value drawn from a normal distribution with mean zero and a standard deviation of $\sigma_{\theta} = 0.1$. We also implement noise in the growth rate and maximum length to prevent the cells from dividing at the same time. The mean of the maximum length corresponds to experimental data \cite{Pin2006,Kiviet2014} and its standard deviation is set to $\sigma_L = 0.1 L_{\text{int}}^{\text{max}}$. Furthermore, daughter cells inherit the growth rate of their mother plus a deviation drawn from a normal distribution with mean zero and standard deviation of $\sigma_\mu = 0.1 \mu_{\text{int}}$. \change{This means that, theoretically, the growth rate can eventually dip below zero. In practice, however, there is not enough time for the growth rates to become that low as we let the colony grow to 4000 particles which takes roughly 12 generations.} The noise parameters are kept at the same value for all our simulations. \change{Because of the noise, we have to measure the mean AR of the eventual simulation. This is done by averaging over all the particles at the end of the simulation where the number of particles is 4000.}

\subsection{Mechanical relaxation}
Our model is based on the actual motion of a growing bacterial colony. On this scale, inertial forces are negligible relative to viscous forces due to the small size of the cells. We therefore expect the system to be in the regime of low Reynolds number, i.e., we consider overdamped dynamics. Thus we neglect inertia and only regard drag to counter applied forces which leads to the following equation of motion for each disk $i$ in our model \cite{VanDrongelen2015}: 
\begin{equation} \label{eq:force}
\bm{F}_{i,\text{tot}}\ =\ \frac{32}{6}\eta D \bm{v}_{i}\ =\ \zeta \bm{v}_{i}
\end{equation}
where $\bm{F}_{i,\text{tot}}$ is the total force acting on the disk, $D$ its diameter, $\bm{v}_{i}$ its velocity and $\eta$ the viscosity of the surrounding material. Here $\zeta$ represents a numerical factor which only depends on the surrounding material since the diameter of the particles is kept constant in our simulations.

During the simulation, growth, division and collision result in a net force acting on each particle. To compute the movement of the particle, this force needs to be converted to a velocity and a displacement over time. Combining the spring forces and the equation of motion results in a force balance. The force balance for disk $i$ is given by:
\begin{equation} \label{eq:forcebalance}
    \bm{F}_{i,\text{int}}\ +\ \bm{F}_{i,\text{ov}}\ =\ \bm{F}_{i,\text{tot}}
\end{equation}
and filling in the forces \change{then} gives
\begin{equation} \label{eq:master}
    -k_{\text{int}}\ \frac{\bm{L}_{i,\text{int}}}{L_{i,\text{int}}}\ (L_{i,\text{int}}^{\text{rest}}\ -\ L_{i,\text{int}})\ +\ \sum_{p \in N_{\text{ov}}} -k_{\text{ov}}\ \bm{\delta}_{p}\ \phi_{ip,\text{ov}}\ =\ \zeta \bm{v}_i
\end{equation}
where $p$ represents each collision that affects disk $i$. At each time step $\Delta t$ the position $\bm{r}$ of disk $i$ is updated according to $\bm{r}_i = \bm{r}_i + \bm{v}_i \Delta t$.

\subsubsection{Characteristic scaling}
To reduce the parameter space of the model we define a characteristic time scale and a unit of length. The only spatial parameter that does not change and has only one value for all elements is their diameter $D$. Therefore, we select the diameter as our unit of length such that all other spatial parameters and measures are scaled with $D$\change{. For the simulations} we set $D$ to be unity, $D = 1$\change{, while for the experiments, we measure an average $D$ to scale the lengths with}. As a result, all length scales are normalized to dimensionless parameters that can easily be compared between different systems (simulations and experiments). To find the characteristic time scale we impose the requirement that for colonies with up to $10^4$ particles no significant overlap (particles cannot cross each other) is possible between particles. This requirement specifies the value of the overlap spring constant $k_{\text{ov}}$.

To derive this value we consider the example with only two disks that overlap. The remaining non-zero elements of Eq. \ref{eq:master} for one of the two disks $i$ gives
\begin{equation} \label{eq:overlap}
   -k_{\text{ov}} \bm{\delta}\ =\ \zeta \bm{v}_i
\end{equation}
where $\bm{\delta}$ is the distance between the centers of the two disks. Since $\bm{\delta}$ is equal to two times the displacement of the disk $\bm{\delta} = 2 \bm{r}_i$ (both disks are moving away from each other in the opposite direction) and the velocity is the time derivative of the position $\bm{v}_i = \dot{\bm{r}}_i$, we can rewrite Eq. \ref{eq:overlap} as
\begin{equation} \label{eq:ode}
    -2 k_{\text{ov}} \bm{r}_i\ =\ \zeta \dot{\bm{r}}_i
\end{equation}
and the solution of this homogeneous first order differential equation is
\begin{equation} \label{eq:sol}
    \bm{r}_i\ =\ c e^{-\frac{t}{\tau}} \quad \text{where} \quad \tau\ =\ \frac{\zeta}{2k_{\text{ov}}}.
\end{equation}
We set $\tau$ as our characteristic time scale and fix the unit of time by setting this time scale to unity, $\tau = 1$. Now we can find the force scale by setting the viscosity term to unity $\zeta = 1$, which leads to the value of the overlap spring constant, $k_{\text{ov}} = 1 / 2 \tau$.

Furthermore, $k_{\text{int}}$ should be in the order of $k_{\text{ov}}$ to keep the particles soft but stiff enough such that even for large colony sizes ($N > 4000$) all particles can still be growing. We therefore set $k_{\text{int}}$ to $k_{\text{int}} = 0.5\ k_{\text{ov}}$.

Growth is an active process and the growth rate does therefore not have a specific dependence on $\tau$. However, the growth rate plays a major role in determining the simulation time and a larger growth rate means a faster simulation. Hence, we prefer the growth rate to be large, but it also needs to meet the requirement of minimal overlap. A large growth rate results in a large amount of overlap since the particles can grow faster than the system is able to relax. To avoid this, the growth rate is related to $k_{\text{ov}}$ and is found empirically. We found the optimal value of the growth rate to be $\mu_{\text{int}} = k_{\text{ov}} \cdot 10^{-4}$. 

\begin{table}[htb]
    \centering
    \begin{tabular}{|l|c|c|}
        \hline
         \textbf{Parameter} & \textbf{Symbol} & \textbf{Value} \\
         \hline
         \textit{Constants} & & \\
         Time scale & $\tau$ & $1$ \\ 
         Diameter & $D$ & $1$ \\ 
         Viscosity term & $\zeta$ & $1$ \\
         \hline
         \textit{Variables} & & \\
         Overlap spring constant & $k_{\text{ov}}$ & $0.5\ \tau$ \\ 
         Internal spring constant & $k_{\text{int}}$ & $0.5\ k_{\text{ov}}$ \\ 
         Growth rate & $\mu_{\text{int}}$ & $1 \cdot 10^{-4}\ k_{\text{ov}}$ \\ 
         Division length & $L_{\text{int}}^{\text{max}}$ & $2\ -\ 8\ D$ \\
         \hline
         \textit{Noise} & & \\
         Growth rate & $\sigma_{\mu}$ & $0.1\ \mu$ \\ 
         Division length & $\sigma_L$ & $0.1\ L_{\text{int}}^{\text{max}}$ \\
         Orientation & $\sigma_{\theta}$ & $0.1$ \\ 
         \hline
         \textit{Observable} & & \\
         Mean aspect ratio & $\phi$ & $\left\langle L_\mathrm{int}/D \right\rangle$ \\ 
         \hline
    \end{tabular}
    \caption{List of parameters used for the simulations.}
    \label{tab:parameters}
\end{table}

\section{Experimental methods}

\change{\subsection{Growing and imaging cells}}
\textit{E. coli} microcolonies were grown in a microfluidic device or on a gel pad on M9 medium supplemented with $0.1\%$ lactose and $0.2\;\mathrm{mM}$ uracil. \change{The bacteria were held onto the surface of the microfluidic device or gelpad by means of a glass coverslip.To be able to perform analysis on large quantities of data, we looked at data from 10 experiments that were conducted initially to probe different scientific questions, from both published \cite{Wehrens2018} and unpublished data. This allowed us to sample a range of different growth conditions, which are} listed in Table~\ref{tab:growthconditions}. We used both wild type and genetically modified \textit{E. coli} from strain MG1655; modified strains had either fluorescent reporter constructs and/or were engineered to express titratable amounts of metabolic enzymes (all used strains are listed in Table~\ref{tab:bacterialstrains}). 

\begin{figure}
\centering
\includegraphics[scale=1]{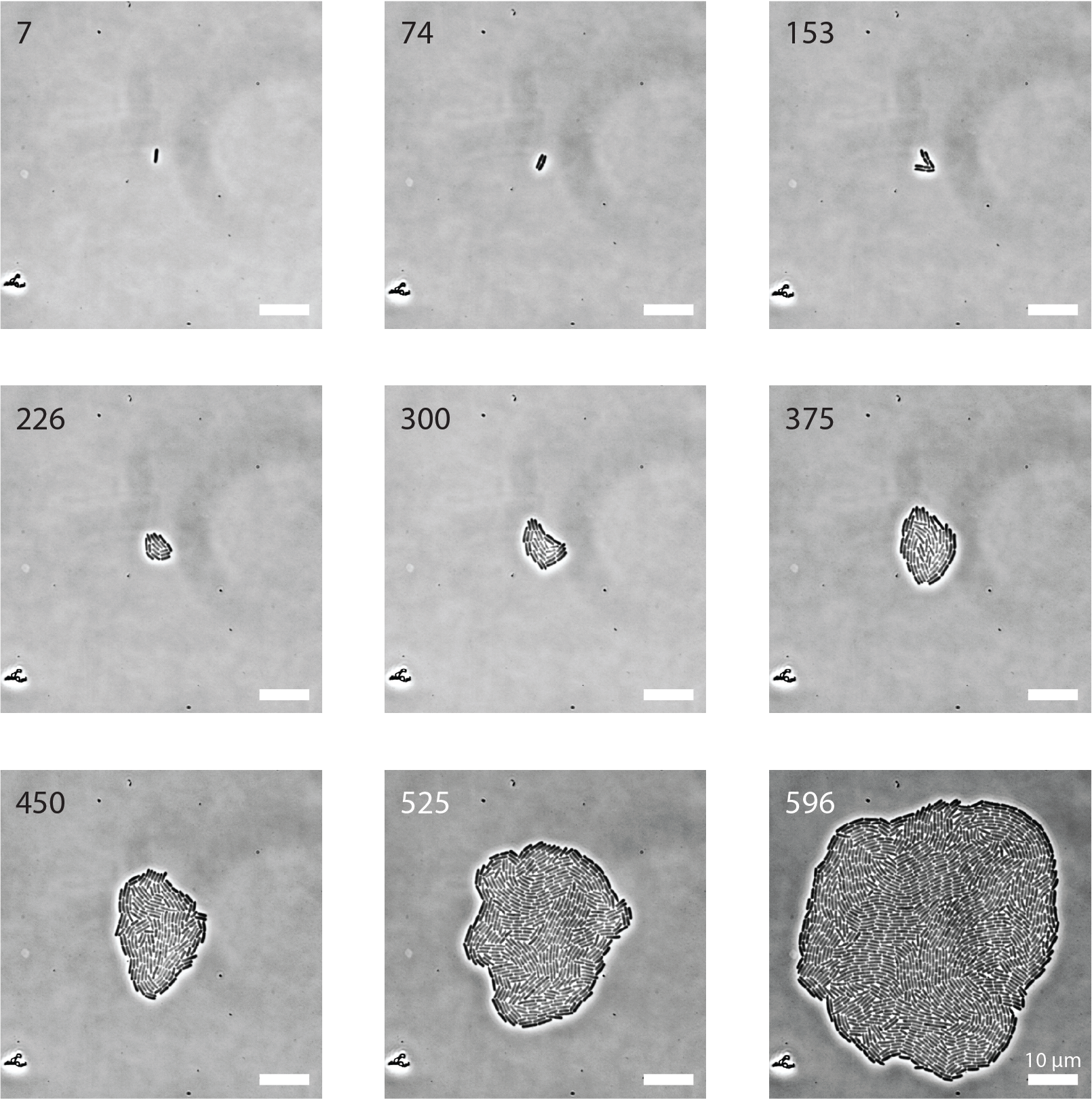}
\caption{Snapshots from the movie of a developing \textit{E. coli} bacterial colony (wild-type). \change{Frame numbers are indicated in the top left corners, recording was done at $1.5$ minutes per frame. Scale bars indicate $10\;\mu\mathrm{m}$. The full movie is available as Supplementary Movie S1.}}
\label{fig:experimentalcolonygrowthsnapshots}
\end{figure}

\begin{figure}
\centering
\includegraphics[scale=1]{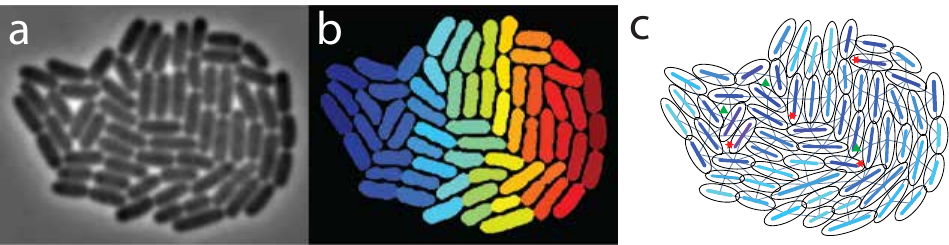}
\caption{\change{Example snapshot of a colony (a), and the resulting segmentation (b), where the colours identify separate particles. In (c) the fitted ellipsoids are shown. The body axis of each ellipsoid is coloured according to the value of the local order parameter (same as in figures~2 and~\ref{fig:particleSOPAR}). Neighbouring bacteria are connected by lines. Defects are shown as red stars ($+\frac12$) and green triangles ($-\frac12$).}}
\label{fig:segmentationsexample}
\end{figure}

\change{Experimental data underlying datapoints 1-3 were previously published~\cite{Wehrens2018}, where the device used is referred to as `microfluidic device 1'. Concisely, \textit{E. coli} in these experiments were pipetted on a microscope cover slip and covered with a 500 {\textmu}m thick polyacrylamide membrane, on top of which a polydimethylsiloxane (PDMS) flow cell was placed that provided a continuous supply of fresh M9 minimal medium to the bacteria through the membrane.
For datapoints 4-10, experiments were conducted on polyacrylamide gel pads. A microscope glass slide with a large rectangular hole was placed on top of a regular microscope glass slide, to create a cavity. Before the experiments, pads of about 5 x 5 mm (with heights identical to the height of the glass slides) were pre-soaked in the desired medium, placed in the cavity, inoculated with bacteria, and covered by a cover slip. The slides were sealed with vacuum grease to prevent evaporation.

For recording videos of the growing colonies, devices or pads were placed under an inverted microscope (Nikon, TE2000), equipped with 100X oil immersion objective (Nikon, Plan Fluor NA 1.3), cooled CMOS camera (Hamamatsu, Orca Flash4.0), xenon lamp with liquid light guide (Sutter, Lambda LS), GFP, mCherry, CFP and YFP filter set (Chroma, 41017, 49008, 49001 and 49003), computer controlled shutters (Sutter, Lambda 10-3 with SmartShutter), automated stage (Märzhäuser, SCAN IM 120 x 100) and an incubation chamber (Solent) allowing precise 37 C temperature control. An additional 1.5X lens was used, resulting in images with pixel size of 0.041 {\textmu}m. The microscope was controlled by MetaMorph software with custom scripts to take images at 1.5-2 min intervals. Fluorescence images were taken in the context of the original experiments, at intervals listed in Table~\ref{tab:growthconditions}. Experiments were stopped when colonies displayed multi-layered growth due to overcrowding.}
\change{
\subsection{Image analysis}}
Images from the experiments were segmented \change{and the particles were tracked in time using a custom MATLAB (MathWorks) program that was originally derived from the \textit{Schnitzcells} package~\cite{Young2012, Wehrens2018}. An example of a segmented dataset can be found in Supplementary video 1. 
The lengths of the particles in each frame were extracted by fitting a polynomial to each curved segmented region along the long axis of the cell and clipping it at the edges. The widths (diameters) of the segmented cells were then calculated by dividing the segmented area by that length. Additionally, the centers of mass were determined and ellipsoids were fitted to the particles to determine their orientation with respect to the x-axis, an example of which can be found in Figure~\ref{fig:segmentationsexample}. 
The aspect ratios of the particles were determined by dividing the length over the width. For each data set an average AR was found by taking the average of all the particles at the point where the colony reaches a size of 500 particles.
The centers and orientations of the particles were used to calculate the orientational order parameters (last frame only) and defect dynamics (whole movies), using the same methods that were used for analyzing the simulation data. To determine the growth rates in each frame, an exponential was fitted over the particle length in the 9 (15 for data set 1) subsequent frames surrounding that frame.} Snapshots from the wild-type movie are shown in Fig.~\ref{fig:experimentalcolonygrowthsnapshots}.

\begin{table}[ht]
\centering
\begin{adjustbox}{width=\textwidth}
\begin{tabular}{|l|l|l|l|l|l|l|}
\hline
\textbf{Data} & \textbf{Bacterial} & \change{\textbf{Aspect}} & \change{\textbf{Symbol}} & \textbf{Additional} & \textbf{Fluorescence reporter,} & \textbf{Device} \\
\textbf{point} & \textbf{strain} & \change{\textbf{Ratio}} & & \textbf{compounds} & \textbf{exposure, and interval$^{***}$} & \\
\hline
1 & ASC555 & 2.9 & $\bigtriangleup$ & $1\;\mu\mathrm{M}$ tetracycline$^*$ & GFP (150 ms), RFP (20 ms) & Microfluidic device 1 \\
&&&&& Interval: 36 min & \\
\hline
2,3& ASC976 & 3.5, 3.6 & $+$ & $0.5\;\mu\mathrm{M}$ tetracycline$^*$ &	 CFP (500 ms), YFP (100 ms) & Microfluidic device 1 \\
&&&&& Interval: 19.5 min & \\
\hline
4,5,6 & ASC1004 & 2.6, 3.0, 3.0 & $\times$ & $800\;\mu\mathrm{M}$ cAMP$^{**}$ & YFP (150 ms), CFP (150 ms) & Acrylamide pad$^{****}$ \\
&&&&& Interval: 16.5 min & \\
\hline
7 & ASC1004 & 1.9 & $\bigtriangledown$ & $80\;\mu\mathrm{M}$ cAMP$^{**}$ & YFP (150 ms), CFP (150 ms) & Acrylamide pad$^{****}$ \\
&&&&& Interval: 30 min & \\
\hline
8 & ASC1004 & 4.4 & $\Diamond$ & $5000\;\mu\mathrm{M}$ cAMP$^{**}$ & YFP (150 ms), CFP (150 ms) & Acrylamide pad$^{****}$ \\
&&&&& Interval: 26 min & \\
\hline
9 & ASC990 & 3.3 & $\bigcirc$ & none &	YFP (150 ms), CFP (150 ms) & Acrylamide pad$^{****}$ \\
&&&&& Interval: 16.5 min & \\
\hline
10 & ASC990 & 3.2 & $\Box$ & $5\;\mathrm{mM}$ aKG & YFP (150 ms), CFP (150 ms) & Acrylamide pad$^{****}$ \\
&&&&& Interval: 16.5 min & \\
\hline
\end{tabular}
\end{adjustbox}
\caption{Strains and conditions per experimental data point. \newline (*) During a defined initial period, bacteria were exposed to a sub-lethal dose of tetracycline, an antibiotic. \newline
(**) cAMP activates CRP, which is a master regulator of metabolic enzyme concentrations. \newline
(***) Cells were exposed to light of wavelengths that excite listed fluorescent reporters (not used in this study).  \newline
(****) Gel pads were pre-soaked in culture media, supplemented with 0.01\% Tween-20.
}
\label{tab:growthconditions}
\end{table}

\begin{table}[ht]
\centering
\begin{tabularx}{0.93\textwidth}{|l|X|l|}
\hline
\textbf{Bacterial strain} & \textbf{Phenotype} & \textbf{Source} \\
\hline
ASC555 & \textit{E. coli}, wild type MG1655 (ilvG- rfb-50 rph-1) & AMOLF\\
\hline
ASC976 & $\Delta$php::pn25-mVenus-cmR,  $\Delta$che::Prrsa-mCerulean-kanR. (Kanamycin and chloramphenicol resistance.) & AMOLF \\
\hline
ASC990 & Wild type strain, except for $\Delta$(galk)::s70-mCerulean-kanR and $\Delta$(intc)::rcrp-mVenus-cmR. (Kanamycin and chloramphenicol resistant.) & AMOLF \\
\hline
ASC1004 & Based on cyaA, cpda null mutant ($\Delta$cyaA  $\Delta$cpda) obtained from Benjamin Towbin, Alon lab (also known as strain bBT80). Introduced $\Delta$(galk)::s70-mCerulean-kanR and $\Delta$(intc)::rcrp-mVenus-cmR. (Kanamycin and chloramphenicol resistant.) & AMOLF / Alon lab~\cite{Towbin2017} \\
\hline
\end{tabularx}
\caption{Bacterial strains used in the experiments}
\label{tab:bacterialstrains}
\end{table}

\clearpage
\section{Additional observations}
\begin{figure}[ht]
	\begin{center}
		\includegraphics[scale=0.7]{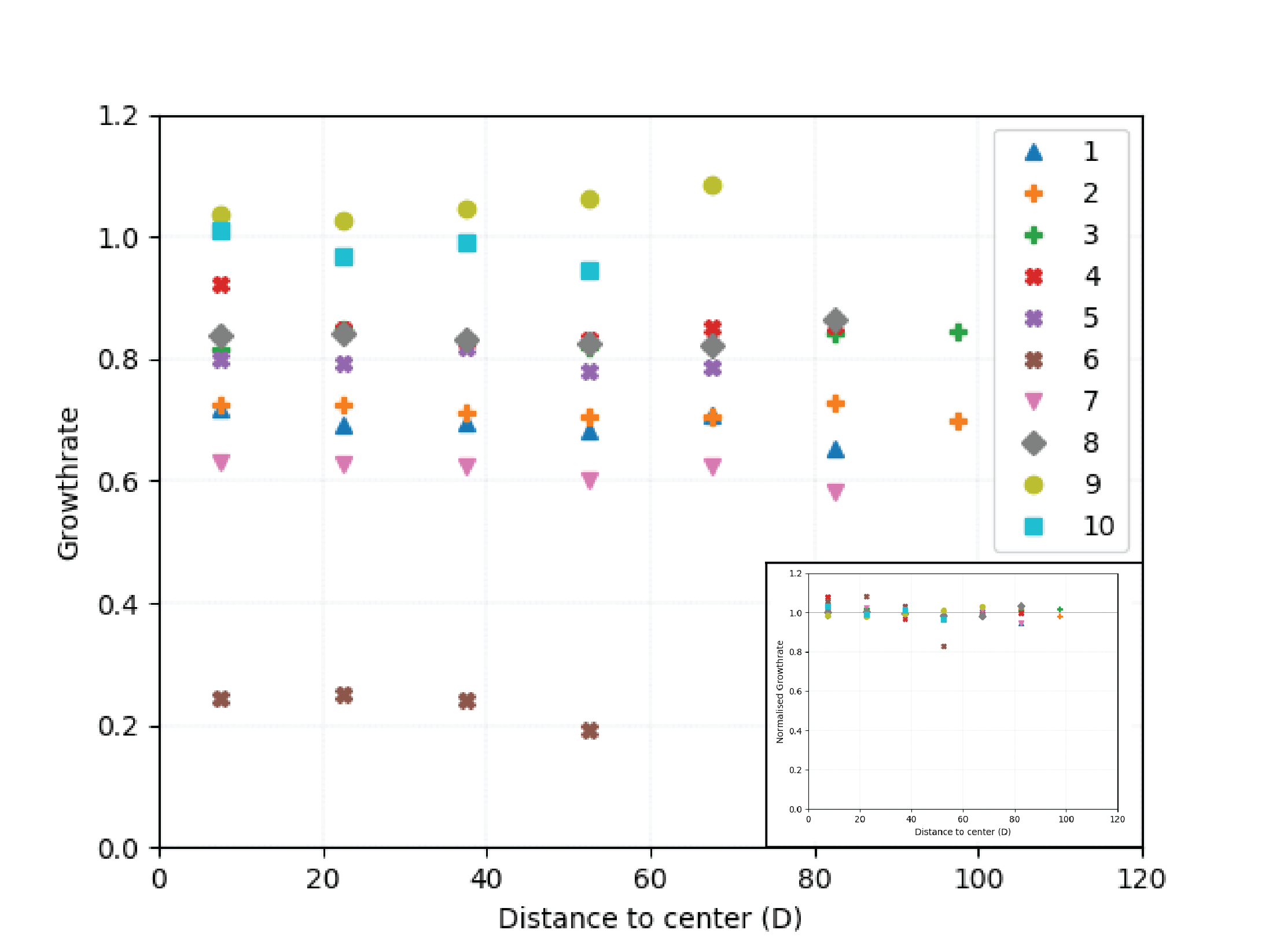}
	\end{center}
	\caption{\change{Fitted growth rates as a function of distance to the center of the tracked colonies. Colony numbers and symbols as mentioned in table \ref{tab:growthconditions}. Growth rates were measured for the duration of an average division cycle (30 frames) at the end of each movie in order to have the largest possible colony size. These growth rates were then binned based on the distance of the centers of the particles to the center of the colony. Note that not all colonies grow to the same number of particles and therefore have different sizes. Inset shows the  normalized growth rate for each colony as a function of radial distance to the center.}}
	\label{growthrates}
\end{figure}

\change{\subsection{Growth rate distribution in experimental colonies}}
\change{As discussed in the methods, we assume a homogeneous growth rate throughout the colony in our simulation. However, it has been shown that this is not always the case for real-life biofilms, as it is possible that bacteria in the center of a colony have limited access to nutrients, or bacteria on the outside are exposed to toxins, for example. To study the growth rate distribution within colonies we have extracted growth rates from the experimental data and plotted them as a function of distance to the center of the colony (Fig. ~\ref{growthrates}). The colonies have various growth rates and sizes, but we clearly see that there is no discernible trend within the colonies. We suspect that this might change if the colonies grow bigger. However, for the early stages of biofilm growth, we consider the assumption of a homogeneous growth rate to be substantiated by these results.}

\begin{figure}[ht]
	\begin{center}
		\includegraphics[scale=1]{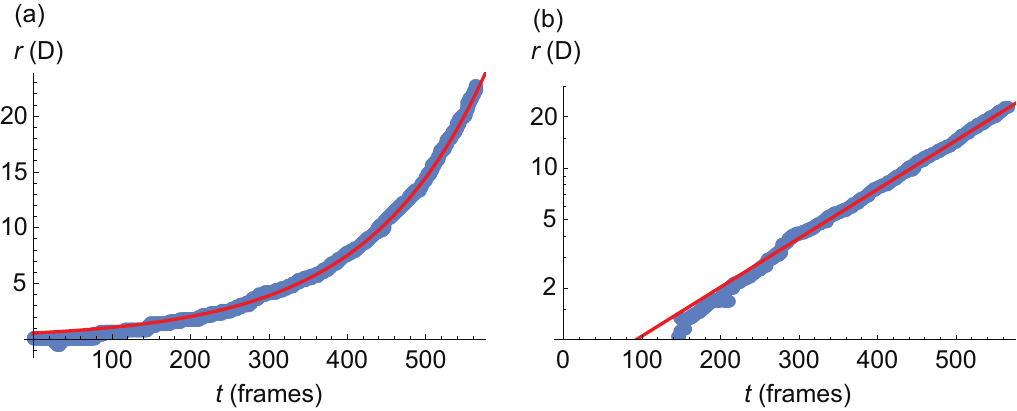}
	\end{center}
	\caption{\change{Radius of experimental colony number 9 as a function of time, on a linear (a) and a log-linear (b) scale. Fitted curves are a single exponential, $r(t) = r_0 \exp(t/\tau)$, with $r_0 = 2.0$ and $\tau = 152$ in this example.}}
	\label{fig:exponentialcolonygrowth}
\end{figure}

\change{The radius of the experimental colonies grows exponentially over time. In figure~\ref{fig:exponentialcolonygrowth} we plot the radius of experimental colony number 9 as a function of time, with a single exponential fit.}

\subsection{Global and local scalar order parameters}
As defined in the main text, the scalar order parameter for a nematic liquid crystal with director~$\textbf{n}$ is given by
\begin{equation}
\label{scalarorderparameter}
S = \langle 2 (\mathbf{w}_i \cdot \mathbf{n}) - 1 \rangle = \langle \cos (2\theta_i) \rangle,
\end{equation}
where $\mathbf{w}_i$ is the orientation of particle $i$, $\theta_i$ the angle between $\mathbf{w}_i$ and $\mathbf{n}$, and the brackets indicate an ensemble average.

\begin{figure}[h]
\centering
\includegraphics[scale=1]{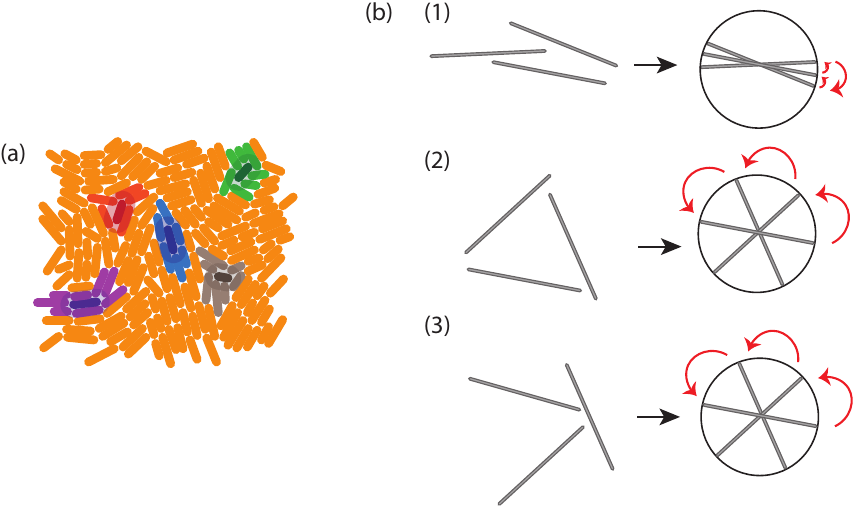}
\caption{\change{Neighbors and defects. (a) Definition of neighboring particles for the calculation of the local order parameter and the position of the defects. The neighbors of a particle are all those particles of which the center of one of their two disks lies within the spherocylinder of diameter $3D$ around the particle of interest. The image shows several such particles (in dark red, blue, green, grey, and purple) with their `catchment area' shaded and all their neighbors in lighter tones of the same color. (b) Algorithm for detecting defects from Zapotocky et al.~\cite{Zapotocky1995}. For any triple of neighboring particles, the particles are positioned on a circle. Starting from the end of arbitrary particle, we then move along the circle to the closest end of another particle, then on to the closest end of the third, and to the closest end of the first. If we return to the same point where we started \change{(case 1)}, there is no defect, but if we end up at the opposite end of the circle \change{(cases 2 and 3)}, there is a defect between the three particles. \change{Depending on the way we go round the defect when we do this procedure, we assign a charge of $-\frac{1}{2}$ (case 2) or $+\frac{1}{2}$ (case 3) to the defect.}}}
\label{fig:neighboranddefectdefinitions}
\end{figure}

\begin{figure}[pht]
\begin{center}
\includegraphics[scale=1]{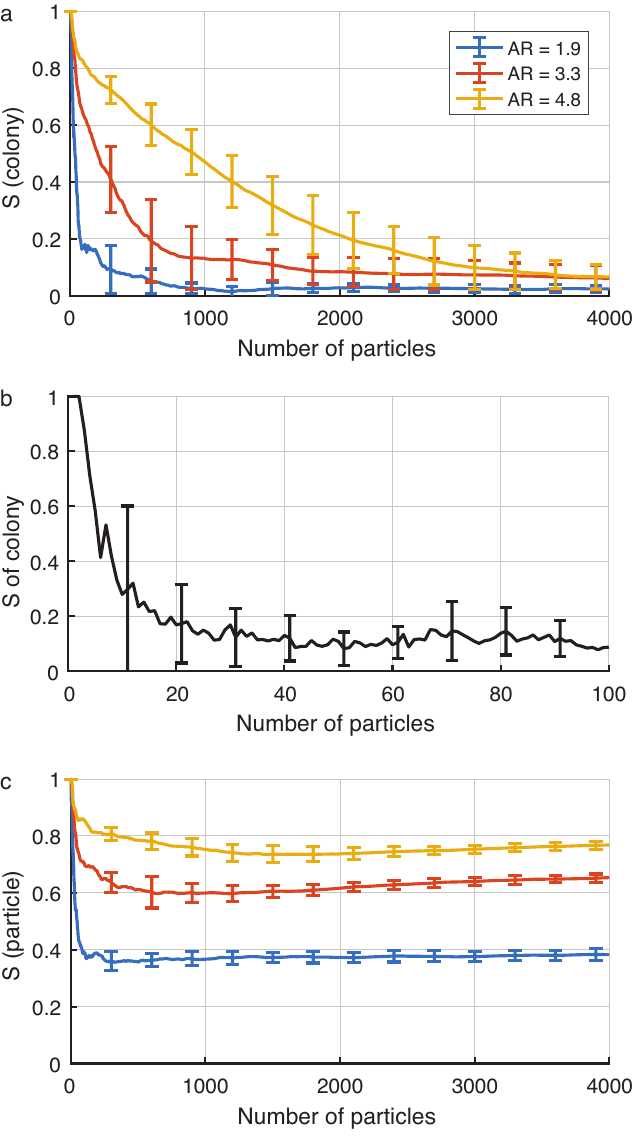}
\end{center}
\caption{Scalar order parameter~$S$ for (a \change{and b}) the entire colony (for simulation and experimental data, respectively) and (c) a particle, as a function of colony size. Colors indicate different particle aspect ratios. \change{Numerical data in (a) are averaged over three runs, error bars show standard deviations. Experimental data in (b) are averaged over eight colonies with mean aspect ratio between 2.9 and 3.5; error bars again show standard deviations. Numerical data in (c) are averaged over all particles in a single simulation; error bars show standard deviations.}}
\label{fig:colonyscalarorderparameter}
\end{figure}

In Figs.~\ref{fig:colonyscalarorderparameter}a \change{and b} we plot the scalar order parameter with the average taken over the entire colony, as a function of colony size, \change{for simulation and experimental data, respectively}. As expected, we find that the value of the order parameter decreases as the number of particles increases, with a rate that depends on the aspect ratio of the particles. In Fig.~\ref{fig:colonyscalarorderparameter}c, we plot the same order parameter where the ensemble consists of the direct neighbors of a particle \change{(defined as illustrated in Fig.~\ref{fig:neighboranddefectdefinitions})}, which we subsequently average over all particles. We find that this value approaches a steady-state as the colony grows. These steady-state values are shown in Fig.~2b of the main text.

\begin{figure}[ht]
\begin{center}
\includegraphics[scale=1]{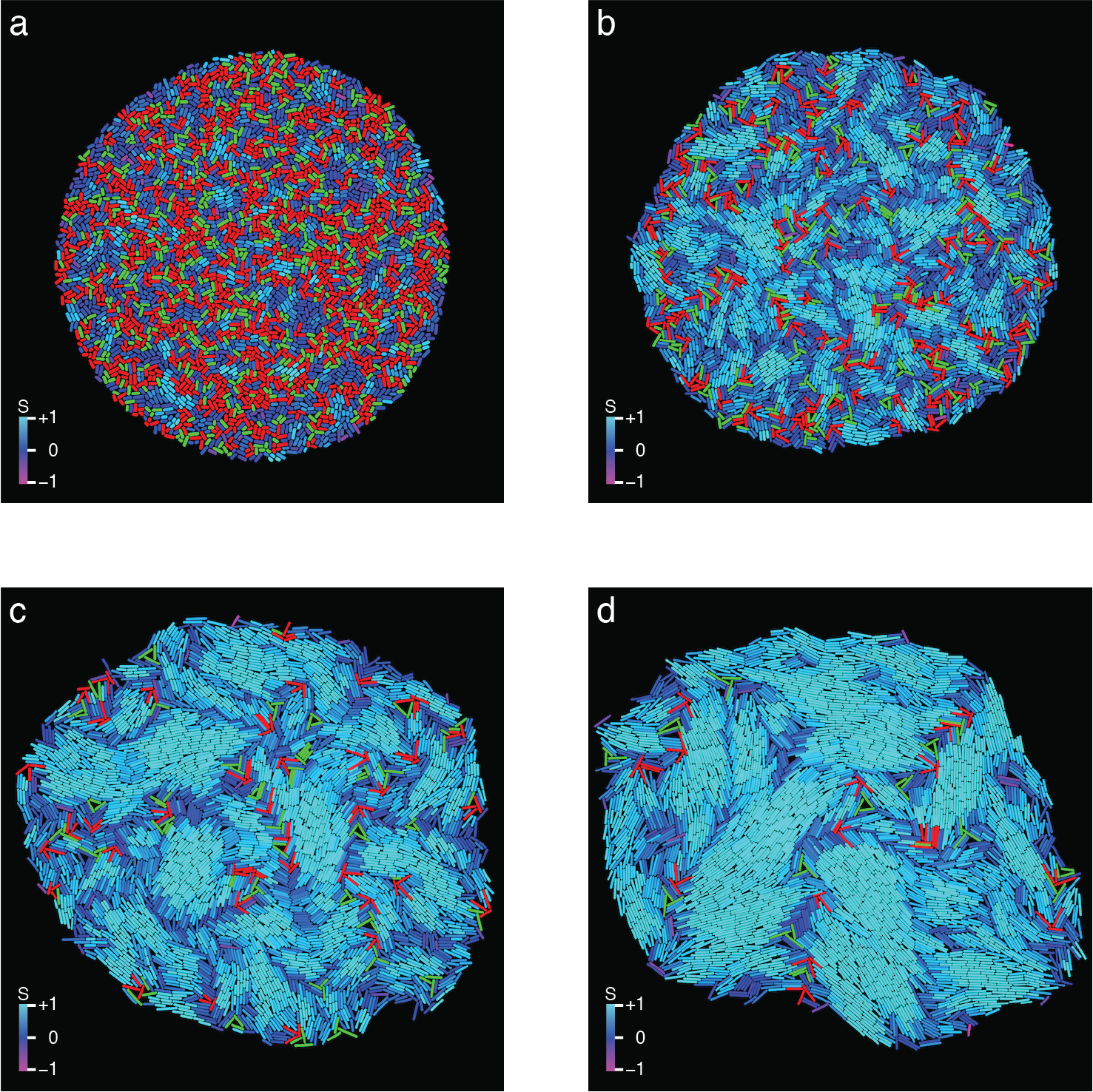}
\end{center}
\caption{Visualization of the scalar order parameter $S_\mathrm{p}$ of individual particles for simulated colonies with different aspect ratios~$\phi$. Light blue equals a high value and \change{purple} a low value of $S_\mathrm{p}$. \change{Red and green denote topological defects of charge $+\frac12$ and $-\frac12$, respectively.} Each snapshot is taken at approximately the same simulation step. (a) $\phi = 1.2$. (b) $\phi = 2.6$. (c) $\phi = 4.0$. (d) $\phi = 5.5$.}
\label{fig:particleSOPAR}
\end{figure}

In Fig.~2a of the main text, we visualized the alignment of the particles by relating the color of the particle to its $S_p$ value (darker blue means a lower value of $S_p$). In this way, domains with a certain orientation are clearly visible. In Fig.~\ref{fig:particleSOPAR} similar images are shown for different aspect ratios.

\subsection{Orientational correlation as a function of distance}
\begin{figure}[ht]
\begin{center}
\includegraphics[scale=1]{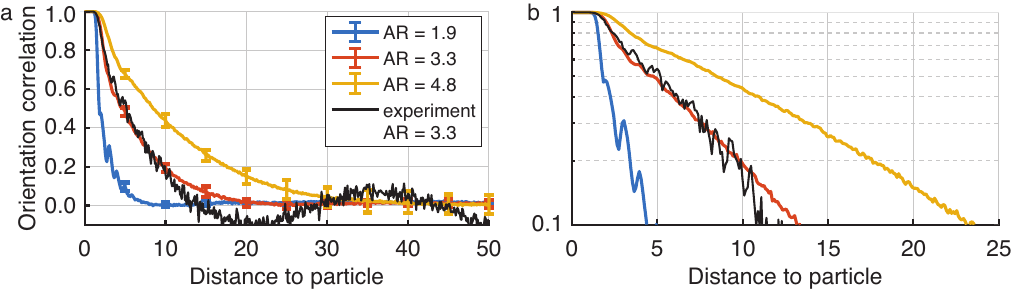}
\end{center}
\caption{Correlation between particle orientation as a function of distance \change{(measured in units of particle diameters)} for three simulated aspect ratios and an experimental dataset, plotted (a) linearly and (b) log-linearly, showing that the correlation decays exponentially. The characteristic correlation lengths of the colonies with different aspect ratios are plotted in Fig.~2c of the main text. \change{Errorbars in (a) represent standard deviations over ten repeats, for clarity only show when the distance to the particle is a multiple of 2.5 particle diameters.}}
\label{fig:correlationdistanceplot}
\end{figure}

\begin{figure}[ht]
\begin{center}
\includegraphics[scale=1]{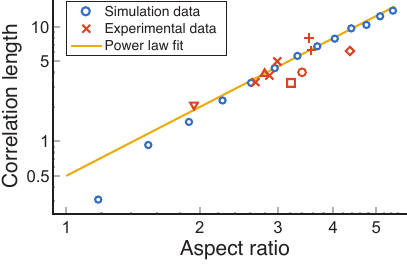}
\end{center}
\caption{\change{Same data as in Fig. 2c, plotted on a log-log scale, showing the correlation length of $S_\mathrm{p}$ versus aspect ratio, for simulation \change{(blue circles)} and experimental \change{(red symbols)} data. The continuous yellow line is a power law fit to the simulation data, with fit parameters $a = 0.50$ and $b = 2.0$.}}
\label{fig:2cinset}
\end{figure}

As Fig.~\ref{fig:particleSOPAR} and Fig.~2a in the main text show, we find orientational domains inside the colony. One way to quantify the size of these domains is by setting a certain maximum threshold value for the difference in orientation between neighbouring particles to distinguish different domains \cite{You2018}. However, as shown in Fig.~\ref{fig:correlationdistanceplot}, there is also an intrinsic length scale associated with the domains, given by the correlation length between the particle orientations. To calculate the correlation as a function of distance, we take a thin ring (with radius $R$ and width $dR$) around \change{the center of} a particle and compute the scalar order parameter of all particles lying in this annulus, again with taking the orientation of the chosen particle as the director. A log-linear plot of this scalar order parameter as a function of distance, Fig.~\ref{fig:correlationdistanceplot}b, clearly shows an exponential decay, which gives us a well-defined correlation length. These correlation lengths are plotted in Fig.~2c of the main text, for both the simulated and the experimental data. \change{A log-log version of Fig.~2c can be found in Fig.\ref{fig:2cinset}}

\begin{figure}[ht]
\centering
\includegraphics[scale=1]{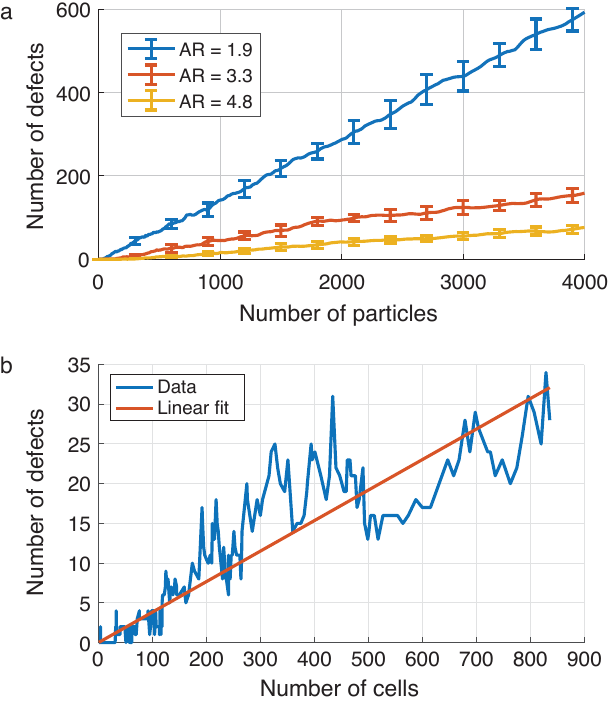}
\caption{\change{Number of defects as a function of the number of particles for (a) simulations and (b) an experimental system. Errorbars in (a) represent standard deviations over ten repeats, for clarity only show when the number of particles is a multiple of 300.}}
\label{fig:defectnumbersplot}

\end{figure}

\change{\subsection{Extracting defect density}
The number of defects in a developing colony scales essentially linearly with the number of particles, as shown in figure~\ref{fig:defectnumbersplot}. The defect densities reported in Fig.~2e of the main text are the slope of the fitted lines.}

\clearpage
\section{Particle density and pressure profile}
\label{sec:density}
\begin{figure}[h!]
\begin{center}
\includegraphics[scale=1]{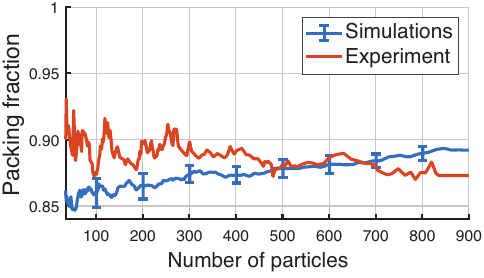}
\end{center}
\caption{The packing fraction of simulations and an experiment with the same aspect ratio ($\phi = 3.3$). \change{Errorbars on the simulation data represent standard deviations over ten repeats, for clarity only shown when the number of particles is a multiple of 100.}}
\label{fig:particledensity}
\end{figure}

A well-established characteristic of bacterial colonies is their particle density, which we  describe with two parameters:
\begin{enumerate}
    \item The packing fraction, $\phi_A$, which corresponds to the fraction of the area of the particles inside the colony with respect to the total area of the colony;
    \item The pressure that particles feel inside the colony.
\end{enumerate}
To compute the packing fraction, we divide the sum of the area of all particles by the total area of the colony. We compute the diameter of each particle as the mean of the distances to its neighbors to avoid overcounting the area due to overlap. The total area $A_c$ is given by the area of the polygon made up by the boundary particles of the colony for $N > 32$. Fig.~\ref{fig:particledensity} shows the packing fraction for the simulations and that of a typical experiment, both with aspect ratio $\phi = 3.3$. In both cases, we find a packing fraction of approximately $0.88$.


The particle density is also related to the amount of overlap between particles. We translate this overlap to the pressure that particle $i$ feels due to the surrounding particles, by taking the sum of the magnitude of the overlap forces $|\bm{F}_{\text{ov}}|$ and scaling with the particle's circumference $C_i$:
\begin{equation}
    P_i = \frac{|\bm{F}_{i,\text{ov}}|}{C_i},\quad \text{where}\quad C_i\ =\ 2 L_{i,\text{int}} + \frac{1}{4} \pi D. 
    \label{eq:pressure}
\end{equation}

The analytic pressure profile in freely growing two-dimensional bacterial colonies will now be derived. To derive this profile, we start from the Stokes equation
\begin{equation}
\label{stokes}
- \nabla p + \eta {\nabla ^2}\mathbf{v} + k \mathbf{v} = \mathbf{f}^{\mathrm{ext}},
\end{equation}
with a boundary condition $p(R)=0$ indicating that the pressure at the edge of the colony is set to zero. Just as in Dell'Arciprete et al., we find a radially symmetric velocity profile in our colonies~\cite{DellArciprete2018}. To solve equation \ref{stokes}, one needs to find an expression for the external force and the velocity. To determine the velocity we start with the continuity equation with a source term often used for growing bacterial colonies~\cite{DellArciprete2018},
\begin{equation}
\frac{{\delta \rho }}{{\delta t}} + \nabla  \cdot (\rho \mathbf{v}) = \gamma \rho,
\end{equation}
in which $\gamma$ is the growth rate. Assuming that density $ \rho$ is constant in both time and space and by assuming a radially dependent velocity profile, $\mathbf{v} = {v_r}\hat r$, we obtain
\begin{equation}
\nabla  \cdot \mathbf{ v} \equiv \frac{1}{r}\frac{d}{{dr}}({v_r}r) = \gamma.
\end{equation}
The solution to this differential equation is
\begin{equation}
{v_r}(r) = \frac{{{c_1}}}{r} + \frac{{\gamma r}}{2}.
\end{equation}
Because the boundary condition ${v_r}(r=0)=0$ needs to hold, $c_1=0$. Thus, for the velocity we obtain
\begin{equation}
\label{velocity}
{v_r}(r) = \frac{{\gamma r}}{2}.
\end{equation}
An interesting result emerges already. It turns out that the radial velocity should only depend on the growth rate and its position to the center of the colony. Using the fact that our problem is radially symmetric the second term in equation \ref{stokes} can be written as
\begin{equation}
\label{vectorlaplacian}
{\eta \nabla ^2}\mathbf{v} \equiv \eta \left( {\frac{1}{r}\frac{\partial }{{\partial r}}\left( {r\frac{{\partial {v_r}}}{{\partial r}}} \right) - \frac{{ {v_r}}}{{{r^2}}}} \right) \hat{r}.
\end{equation}
Equation \ref{velocity} can be substituted in equation \ref{vectorlaplacian} to obtain
\begin{equation}
\label{vectorlaplacianresult}
\eta {\nabla^2}\mathbf{v} = \eta \left( {\frac{\gamma}{{2r}} - \frac{\gamma}{{2r}}} \right) \hat{r} = 0.
\end{equation}
The next step is to find a proper description for the force. The only force present in the system emerges from particle growth. To compensate for the fact that an increase in radius results in more particles being present, we scale this active force with $r/R$. In this way, the force of growing is the largest when all particles are included. By doing so and using the result obtained in equation~\ref{vectorlaplacianresult}, equation~\ref{stokes} simplifies to
\begin{equation}
\frac{{\partial p}}{{\partial r}} = \left(\frac{{{f^{\mathrm{act}}}}}{R}+\frac{k \gamma}{2} \right) r.
\end{equation}
Using the boundary condition $p(R)=0$, the differential equation is easily solved as
\begin{equation}
\label{finalpressure}
p(r) =  \left(\frac{{{f^{\mathrm{act}}}}}{2R}+\frac{k \gamma}{4} \right) \left( {{R^2} - {r^2}} \right).
\end{equation}
The analytic result indeed shows a quadratic pressure profile, which is also observed in our simulations (cf. Fig.~1b of the main text). 

\begin{figure}[ht]
	\begin{center}
		\includegraphics[scale=0.7]{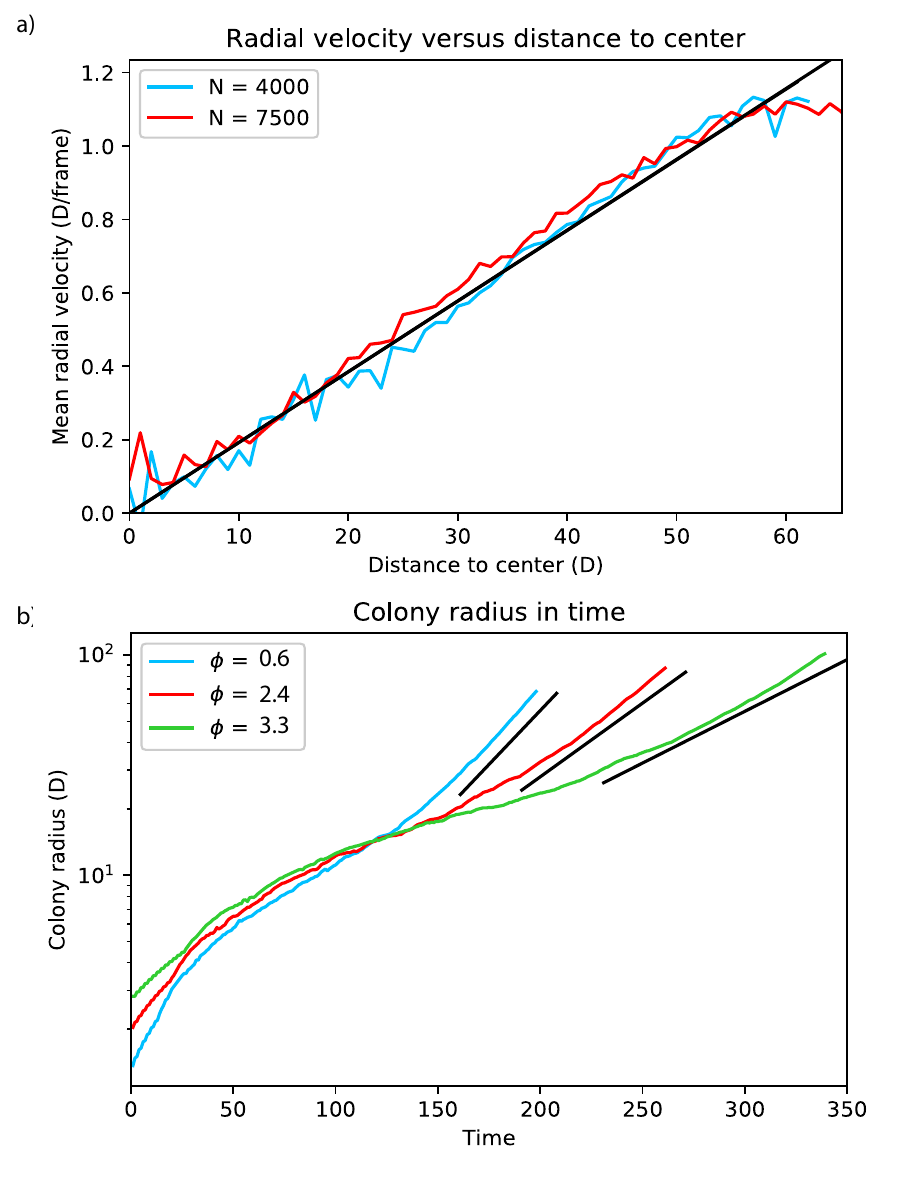}
	\end{center}
	\caption{(a) The velocity is a linear function of the distance to the center \change{measured in particle diameter~($D$)}. The slope is given by $\frac12 \alpha \gamma$, where $\gamma$ is the growth rate and $\alpha$ a correction factor for particles with aspect ratios larger than zero. The slope is constant in time. \change{Black line: linear fit with slope~$1.54 \cdot 10^{-2}\;\mathrm{(frame)}^{-1}$}. (b) The logarithm of radius grows linear in time once the colony has become roughly circular for different aspect ratios. The slope is given by $\frac12 \beta \gamma$, with $\gamma$ again the growth rate and $\beta$ another correction factor. \change{Black lines represent slopes of (left to right) $9.8 \cdot 10^{-3}$, $9.1 \cdot 10^{-3}$ and $6.3 \cdot 10^{-3}$ per unit time.}}
	\label{velocityandradius}
\end{figure}

The simulated data shows that normalized pressure versus normalized distance squared to colony center collapses to a single line for different times. This relation is described by
\begin{equation}
\label{normpressvsnormrad}
\frac{{P(r,t)}}{{P(r = 0,t)}} = 1 - \frac{{{r^2}}}{{{R^2}(t)}}.
\end{equation}
From equation \ref{finalpressure}, $P(r=0,t)$ can be obtained. By doing so, equation \ref{normpressvsnormrad} can be rewritten as
\begin{equation}
P(r,t) = \left( \frac{1}{2}{f^{\mathrm{act}}}R(t) + \frac{k \gamma}{4}R^2(t)  \right) \left( {1 - \frac{{{r^2}}}{{{R^2}(t)}}} \right).
\end{equation}
Using equation \ref{velocity},  a relation for $R(t+\Delta t)$ can be obtained
\begin{equation}
R(t + \Delta t) - R(t) = \frac{1}{2}\gamma R(t)\Delta t.
\end{equation}
By taking the limit $\Delta t \to 0$ and solving the differential equation, we obtain
\begin{equation}
R(t) = {R_0}{e^{\frac{1}{2}\gamma t}}.
\end{equation}
Substituting this result into equation \ref{finalpressure}, results in the pressure profile 
\begin{equation}
\label{fullpressure}
P(r,t) = \left( \frac{{{f^{\mathrm{act}}}}}{{2{R_0}}}{e^{ - \frac12\gamma t}} + \frac{k \gamma}{4} \right) \left( {R_0^2{e^{\gamma t}} - {r^2}} \right).
\end{equation}
From this, we see that the pressure in the center blows up exponentially when time evolves. As can be seen in Fig.~\ref{velocityandradius}, we indeed find a linear radial velocity and an exponentially increasing colony radius in our simulations. However, a deviation from the simple hydrodynamic model is observed in both cases. To capture these deviations, two correction factors, $\alpha$ and $\beta$, can be introduced. Since the velocity stays linear, it will only affect the magnitude of the pressure. The correction factor for radius growth does affect the behavior of the pressure in time, but not in space. Therefore, we still find a quadratically decaying spatial pressure profile in simulations. The correction factors can be included to obtain the in simulations observed pressure profile
\begin{equation}
P(r,t) = \left(    \frac{{{f^{\mathrm{act}}}}}{{2{R_0}}}{e^{ - \frac12\beta \gamma t}} + \frac{\alpha k \gamma}{4}\right) \left( {R_0^2{e^{\beta \gamma t}} - {r^2}} \right). 
\end{equation}

\change{
\section{Supplementary movies}
\noindent\textbf{Movie~S1}: Recording of the growth of one of our experimental colonies. Recording at 1.5 minutes per frame. Snapshots of the movie with scale bars are given in Fig.~\ref{fig:experimentalcolonygrowthsnapshots}.\\

\noindent\textbf{Movie~S2}: Visualization of a simulation of a growing bacterial colony. Bacteria are colored according to the same color scheme as used in Fig.~2(d) of the main text.
}

\section{Author contributions}
\noindent R. L. analyzed data, performed simulations, revised the manuscript.\\
D. v. H. developed and performed simulations, analyzed data, wrote the manuscript.\\
G. N. performed simulations, analyzed data, wrote the manuscript.\\
M. W. performed experiments, analyzed data.\\
S.J. T. designed research, supervised experiments.\\
T. I. designed research, supervised simulations, wrote the manuscript.\\
\clearpage

%

\end{document}